\documentclass[
    ,final            
  ]
  {aipproc}

\layoutstyle{8x11double}

\def\na{\mathcal{N}^2(\phi)}
\def\pe{\phi_J'}
\def\pj{\phi_{J}^{1}}

\begin{document}

\title{Density anomalies and high-order jamming crossovers}


\author{Massimo Pica Ciamarra}{
  address={CNR--SPIN, Dip. di Scienze Fisiche, Univ. di Napoli Federico II,
I-80126, Napoli, Italy}
}

\author{Peter Sollich}{
  address={King's College London, Department of Mathematics, Strand, London WC2R
2LS, United Kingdom}
}

\begin{abstract}
Jamming crossovers occur at zero temperature in assemblies of particles interacting via finite
range repulsive potentials, when on increasing the density particles make contacts with those of subsequent coordination shells.
Density anomalies, including an increased diffusivity upon isothermal compression and a negative thermal expansion
coefficient, are the finite temperature signatures of these jamming crossovers. In this manuscript we show 
that the jamming crossovers are correlated with an increase in the non-affine response of the system to density changes,
and demonstrate that jammed systems evolve upon compression through
successive avalanches triggered by plastic instabilities.
\end{abstract}

\maketitle


\section{Introduction}
On increasing the density of a liquid, particles find themselves in a more
crowded environment. Most frequently, crowding inhibits particle motion and
reduces diffusivity. However, this is not always the case, as there are
liquids whose diffusivity increases upon isothermal compression
in some range of control parameters. This phenomenon is known as a density anomaly. 
Liquids with this anomaly are also frequently characterized, 
in or near the same region of the phase diagram, 
by another anomaly, namely, a negative thermal expansion coefficient. 
Despite their names, density anomalies are quite common. 
The most noticeable anomalous liquid is water, but other examples include Si, Ge,
Sn, and ionic melts with suitable radius ratio, such as SiO$_2$, BeF$_2$ and GeO$_2$~\cite{Evans1999,Miller2009}. 

We have recently shown that in systems of particles 
interacting via finite ranged purely repulsive potentials,
density anomalies are the finite temperature counterpart
of high order jamming crossovers~\cite{PCS1,PCS2}.
Jamming crossovers are conveniently described by considering
the spatial structure of disordered particulate assemblies 
in terms of a series of coordination shells. 
At zero temperature, the jamming transition occurs on increasing the density
when particles cannot avoid forming contacts with their neighbours in
the first coordination shell~\cite{vanHecke2010,Liu2010}. On further increasing the density, successive
jamming crossovers occur as particles are forced to make contacts with neighbours
in the following coordination shells. 

In this paper, after a short review of the main features of the jamming crossovers,
we focus on the correlation between jamming crossovers and non-affinity.
We show that the crossovers induce an increase in the non-affine response, 
and also clarify that this is correlated to a compression-induced avalanche dynamics. 
Avalanches give rise to a long ranged displacement field, and to a logarithmic
scaling of non-affinity with system size.

\section{High--order jamming crossovers}
\subsection{Introduction}
Systems of particles interacting via repulsive contact forces undergo
a jamming transition when their volume fraction crosses a
threshold $\phi_J$. See~\cite{vanHecke2010,Liu2010} for recent reviews.
At the jamming transition each particle is forced to make contacts
with some neighbours, and the mean contact number, which is $Z = 0$ below the transition,
jumps to the isostatic value $Z_{\rm iso}$. The isostatic contact
number, $Z_{\rm iso} = 2d$
in $d$ spatial dimensions, is the minimum numer of contacts required 
for mechanical stability, according to Maxwell counting~\cite{Maxwell}.
Above the jamming transition the bulk and the shear modulus of the system
scale as powers of the excess contact number, $\Delta Z = Z-Z_{\rm
  iso}$, which numerical results suggest grows as $\Delta Z \propto (\phi-\phi_J)^{1/2}$.
The excess contact number also identifies a diverging length scale into the system.
Indeed, by removing all of the bonds of a blob of radius $l$, the excess contact
number of the blob decreases, and vanishes when the blob size equals
a length scaling as $\xi~\propto 1/\Delta Z$, which diverges at the transition.
This length influences the density of 
vibrational modes of the system, which satisfies the Debye scaling $D(\omega) \propto \omega^{d-1}$
up to a characteristic frequency $\omega^*(\phi) \propto \Delta Z(\phi)$. The analysis of
the spatial features of the eigenvectors corresponding to this characteristic 
frequency reveals the presence of a correlation length scaling as $\xi$~\cite{Silbert2009}.
Scaling relations in $\phi-\phi_J$ hold for the geometric and mechanical properties of jammed
systems in the proximity of the jamming transition, i.e.\ in the limit $\phi \to \phi_J$.

We have recently investigated how the above quantities behave at high density,
well above that of the jamming transition, by performing molecular dynamics simulations
of a $50:50$ bidisperse mixture of $N$ disks of diameter $D_l$ and $D_s = D_l/1.4$.
Two particles of average diameter $D$, at a distance $r$, interact when in contact,
$\delta = D-r > 0$, via a potential 
\begin{equation}
V(\delta) = \frac{1}{\alpha}\left(\frac{\delta}{D_l}\right)^\alpha.
\end{equation}
The parameter $\alpha$ sets
the softness of the interaction. The larger $\alpha$, the softer the
interaction, given that $\delta/D_l < 1$.
We have investigated different values of $\alpha$ and of the number of particles $N$.
For each set of parameters, we have minimized the energy of the system via the conjugate gradient
method, starting from random particle configurations. 

Here we review briefly the geometric and mechanical features of the jamming crossovers of a
two dimensional system with $\alpha = 2$~\cite{PCS1,PCS2}, and then analyze 
how the jamming crossovers influence the non-affinity of the system.

\subsection{Jamming crossovers at zero temperature}
\subsubsection{Geometric properties}
\begin{figure}[t!!]
\includegraphics*[width=1\linewidth]{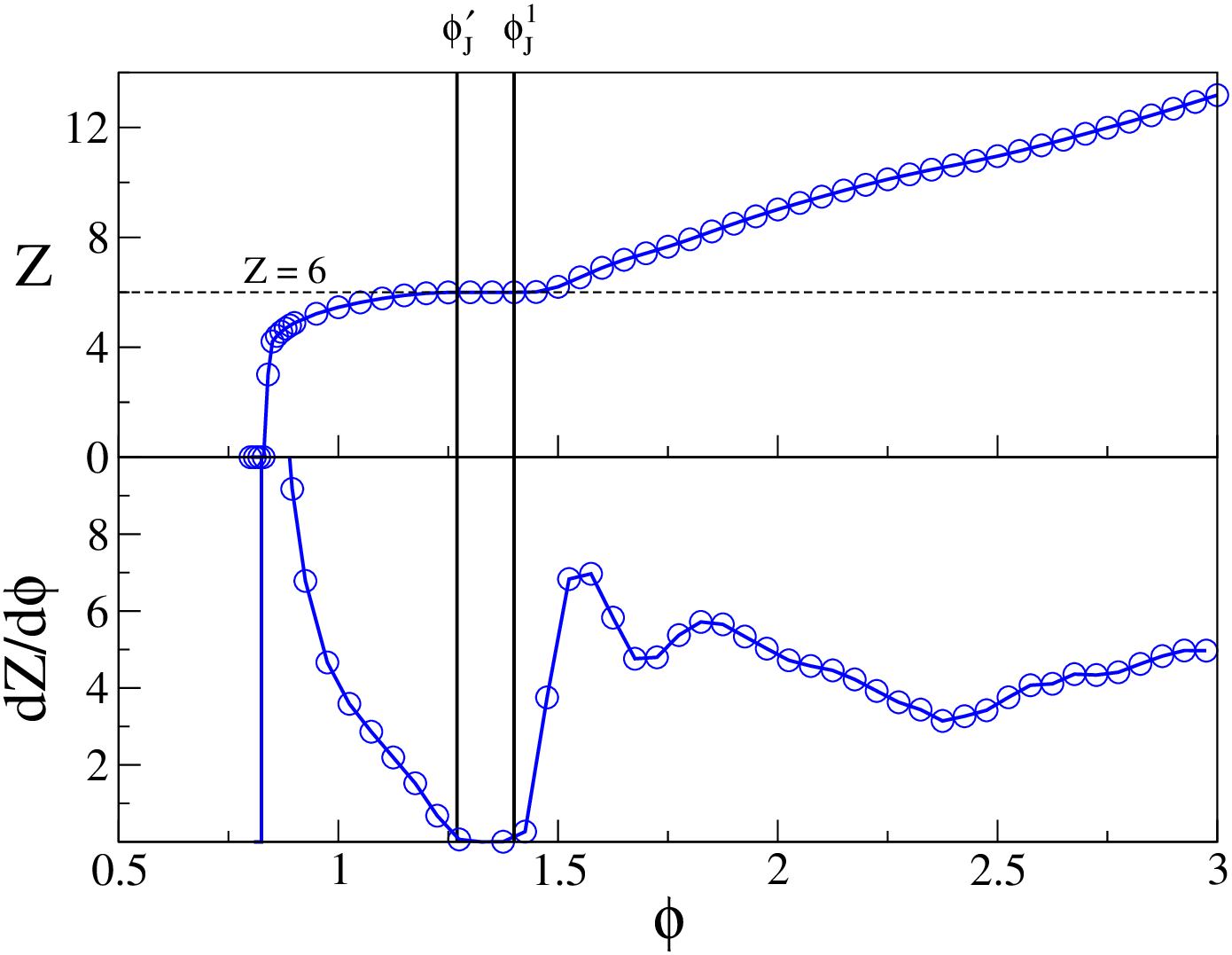}
\caption{
\label{fig:Z}
Volume fraction dependence of (a) the mean contact number $Z$, and (b) its volume
fraction derivative $dZ/d\phi$, for a system of $N = 10^4$ particles with $\alpha = 2$.}
\end{figure}

Fig.~\ref{fig:Z} illustrates the volume fraction dependence of the average contact number and of its
volume fraction derivative. After reaching the isostatic value at the jamming transition,
$Z$ grows monotonically with the volume fraction.
The rate of formation of new contacts, $dZ/d\phi$, oscillates on compression,
and is related to the oscillations of the value of the radial
distribution function at contact.

Fig.~\ref{fig:Z} shows that there is a volume fraction range in which the average contact number is 
constant and equals $Z = 6$. This volume fraction range extends from a volume fraction $\pe$, 
up to the volume fraction of the first jamming crossover $\pj$.
At the volume fraction $\pe$, the formation of contacts with particles in the first coordination shell ends.
In the volume fraction range $\pe$--$\pj$, no contacts are formed or
destroyed, and $dZ/d\phi = 0$ as can be seen in Fig.~\ref{fig:Z}.
The formation of contacts with particles in the second coordination
shell then begins at $\pj$.

The radical Voronoi tessellation~\cite{Radical} of the system reveals that in 
the volume fraction range $\pe$--$\pj$ two particles are in 
contact if and only if they are Voronoi neighbors. This allows
one to rationalize the value $Z = 6$ of the contact number using
Euler's theorem for planar graphs, which fixes to $6$ the average connectivity of any tessellation of space in two dimensions.
Consequently, $\pe$ is the volume fraction at which the
fraction $\theta_1$ of Voronoi neighbors that are not in contact vanishes on increasing the density,
while $\pj$ is the volume fraction at which the fraction of contacts between particles which are not Voronoi neighbors vanishes
on decreasing the density. Numerical results~\cite{PCS2} give the estimates $\pe \simeq 1.27$ and $\pj \simeq 1.40$.

\begin{figure}[t!!]
\includegraphics*[scale=0.33]{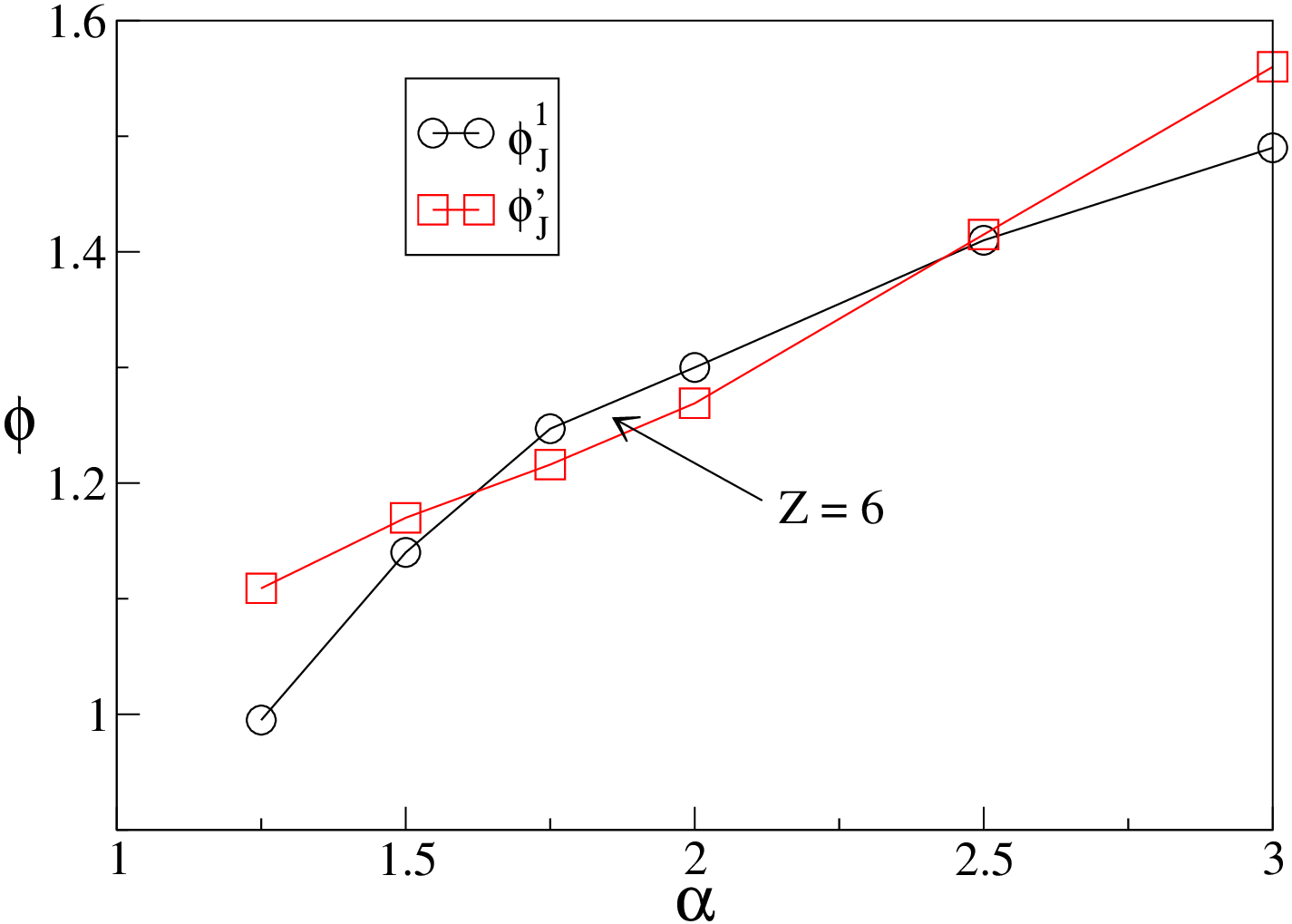}
\caption{
\label{fig:jpj1}
Dependence of the volume fractions $\pe$ (cricles) and $\pj$ (square) on the softness of the interaction 
potential $\alpha$. When $\phi$ reaches $\pe$, all Voronoi neighbors become contacting particles.
When $\phi$ overcomes $\pj$, contact between particles which are not Voronoi neighbors emerge.
}
\end{figure}

The relation $\pe < \pj$ is not verified in three dimensions, where
also Euler's theorem does not constrain  $Z$.
In addition, its validity also depends on the interaction potential, which affects the radial
distribution function and hence the separation between the first and the second coordination
shell. We have explicitly investigated the $\alpha$ dependence 
of $\pe$ and $\pj$, and show in Fig.~\ref{fig:jpj1} that
there is a finite range of softness of the interaction potential in which $\pe < \pj$,
with $Z = 6$ between these volume fractions. Outside this range of
$\alpha$, contacts with first neighbours are still being formed when
the process of making contacts with neighbours in the second
coordination shell begins. 

\subsubsection{Mechanical properties}
\begin{figure}[t!!]
\includegraphics*[scale=0.3]{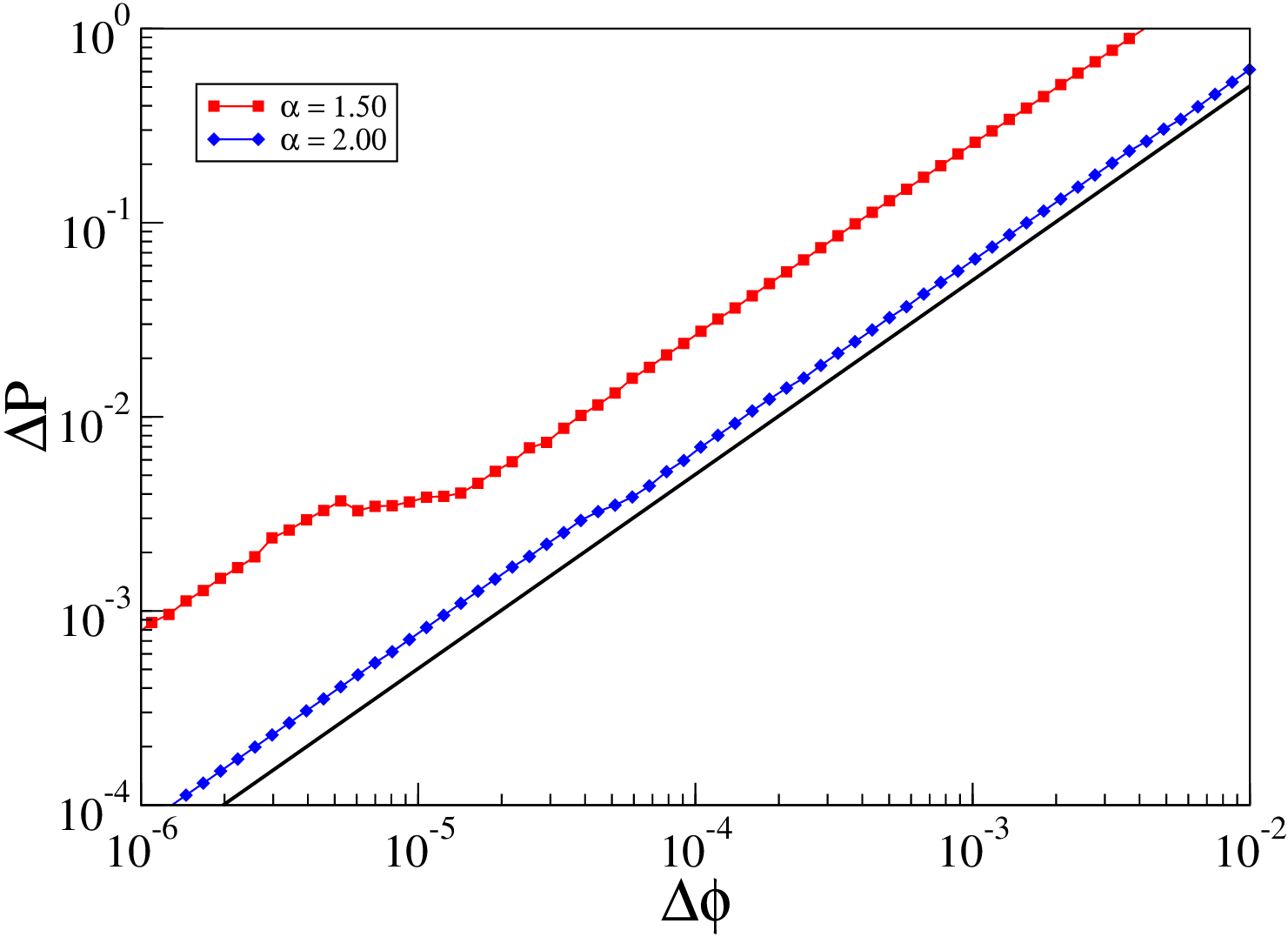}
\caption{
\label{fig:DP}
Volume fraction dependence of the pressure increment resulting from a
quasistatic volume fraction increase by $\Delta \phi$, starting from
$\phi = 0.9$. We define the bulk modulus from the slope of the $\Delta
P(\Delta\phi)$ relation at the upper end of the range of $\Delta \phi$.
}
\end{figure}

\begin{figure}[t!!]
\includegraphics*[scale=0.33]{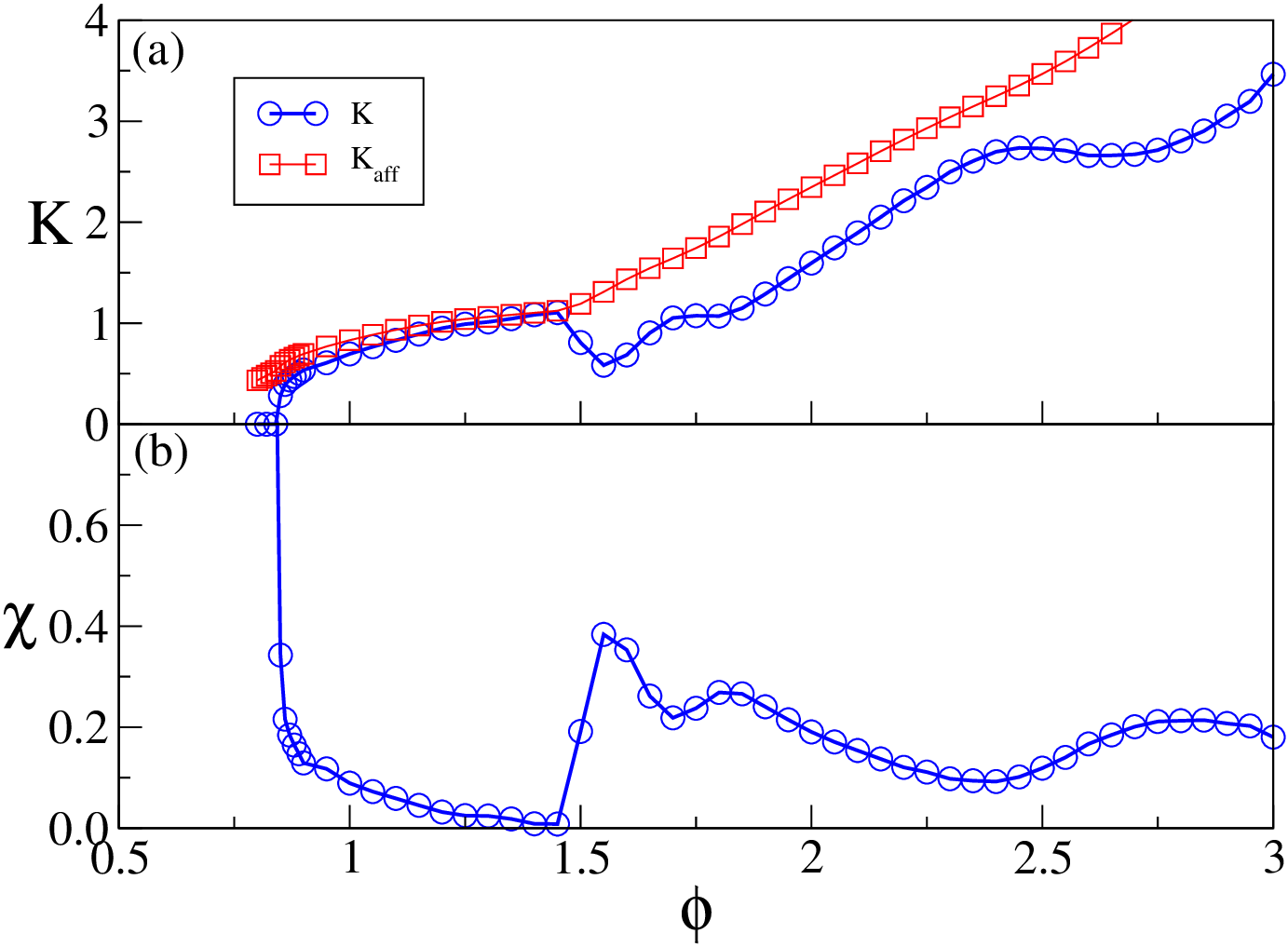}
\caption{
\label{fig:mec}
Volume fraction dependence of (a) the bulk modulus $K$, and of (b) the non affine parameter $\chi$.
$K$ decreases and $\chi$ increases around the jamming crossovers.
}
\end{figure}

The jamming crossovers induce changes in the mechanical properties of the system.
For $\alpha = 2$, the crossovers weaken the system, as illustrated 
by the volume fraction dependence of the bulk modulus, $K = VdP/dV = \phi dP/d\phi$. 
In order to compute $K$, we have quasistatically increased
the volume fraction by an amount $\Delta\phi$, and monitored the corresponding pressure change $\Delta P$.
As an example, we show in Fig.~\ref{fig:DP} the dependence of $\Delta P$ on $\Delta \phi$ for $\phi = 0.9$, for two values of $\alpha$.
The figure clarifies that $\Delta P$ increases linearly with $\Delta \phi$ both at very small $\Delta \phi$,
as well as at larger $\Delta \phi$. These two linear regimes are connected by a range of volume fraction in which
$\Delta P$ grows slowly with $\Delta \phi$. The location of these crossover regions, which originates
from compression induced avalanches as discussed in the next section, depend on both the volume
fraction and the softness of the interaction potential. 
Here we have defined the bulk modulus taking for
$dP/d\phi$ the slope of the $\Delta P(\Delta\phi)$ relation found at
not-too-small incremental compressions, $\Delta \phi = 10^{-2}$. 
The resulting dependence of the modulus on the volume fraction is illustrated in Fig.~\ref{fig:mec}a. 
The figure clarifies that the there exist volume fraction ranges in which the bulk modulus decreases on compression. 
These volume fraction ranges are correlated
with those where the mean contact number increases upon compression, i.e.\ with the jamming crossovers.

We also show in Fig.~\ref{fig:mec}a the bulk modulus calculated in the affine (Born) approximation. 
In this approximation, the negative contribution to the modulus
due to the fluctuation term of the stress tensor~\cite{fluctuation_modulus} is neglected, so that 
$K_{\rm aff} \ge K$. The comparison of the two moduli allows one to define the following
non-affinity parameter,
\[
 \chi = \frac{K_{\rm aff}-K}{K_{\rm aff}+K}, \qquad 0 \leq \chi \leq 1.
\]
$\chi = 0$ when the response is affine, while $\chi \to 1$ when the response is highly non--affine.
Fig.~\ref{fig:mec}b illustrates the volume fraction dependence of $\chi$, and clarifies that
the degree of non-affinity increases on approaching both the jamming transition
and the high order jamming crossovers. Consistent with this, the density
of states reveals the emergence of an abundance of soft modes in these
regimes~\cite{PCS1,PCS2}.

\section{Non--affine response}
\begin{figure}[t!!]
\includegraphics*[scale = 0.3]{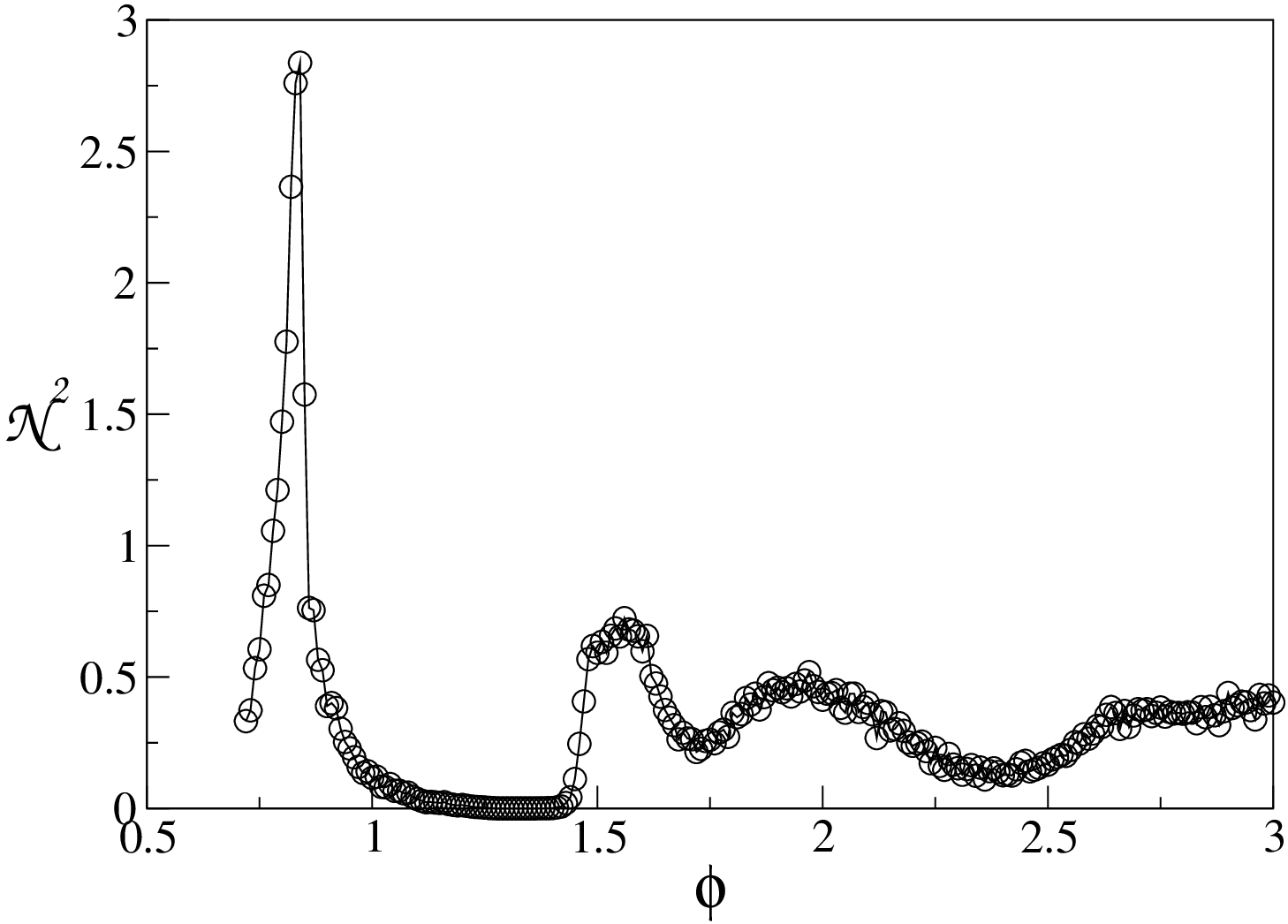}
\caption{
\label{fig:affine1000}
Volume fraction dependence of the non-affine parameter $\na$, for different
a system of $N = 10^3$ particles, and $\alpha = 2$.}
\end{figure}

A direct measurement of non-affinity can be obtained by tracking the displacement of 
the particles on compression. 
To this end, after imposing a small affine compression on a jammed system, we
minimize the energy and measure the non-affine displacement of the particles.
Specifically, we have varied the volume fraction by inflating the diameters of the particles at constant
volume, so that the actual displacement of the particles is the non-affine field.
We quantify the magnitude of this displacement field via the parameter
\begin{equation}
\na = \frac{1}{N} \frac{\sum_i \left[ {\bf r}_i(\phi+\Delta\phi) - {\bf r}_i(\phi) \right]^2 }{L^2 (\Delta \phi)^2},
\end{equation}
where $L$ is the system size, and have calculated it for $\Delta \phi = 10^{-2}$, for different system sizes. Note that up to a factor $(1+\Delta\phi/\phi)$, $\na$ is identical to the mean-squared difference between non-affine and affine particle displacements when the system is compressed and particle sizes kept fixed. The 
normalization factor $L^2 (\Delta \phi)^2$ is chosen as it gives the typical size of the mean-square affine displacements under such a compression.
The numerical results in Fig.~\ref{fig:affine1000} confirm that the jamming crossovers
induce an increase of the non-affinity of the system.

\begin{figure}[t!!]
\includegraphics*[scale = 0.3]{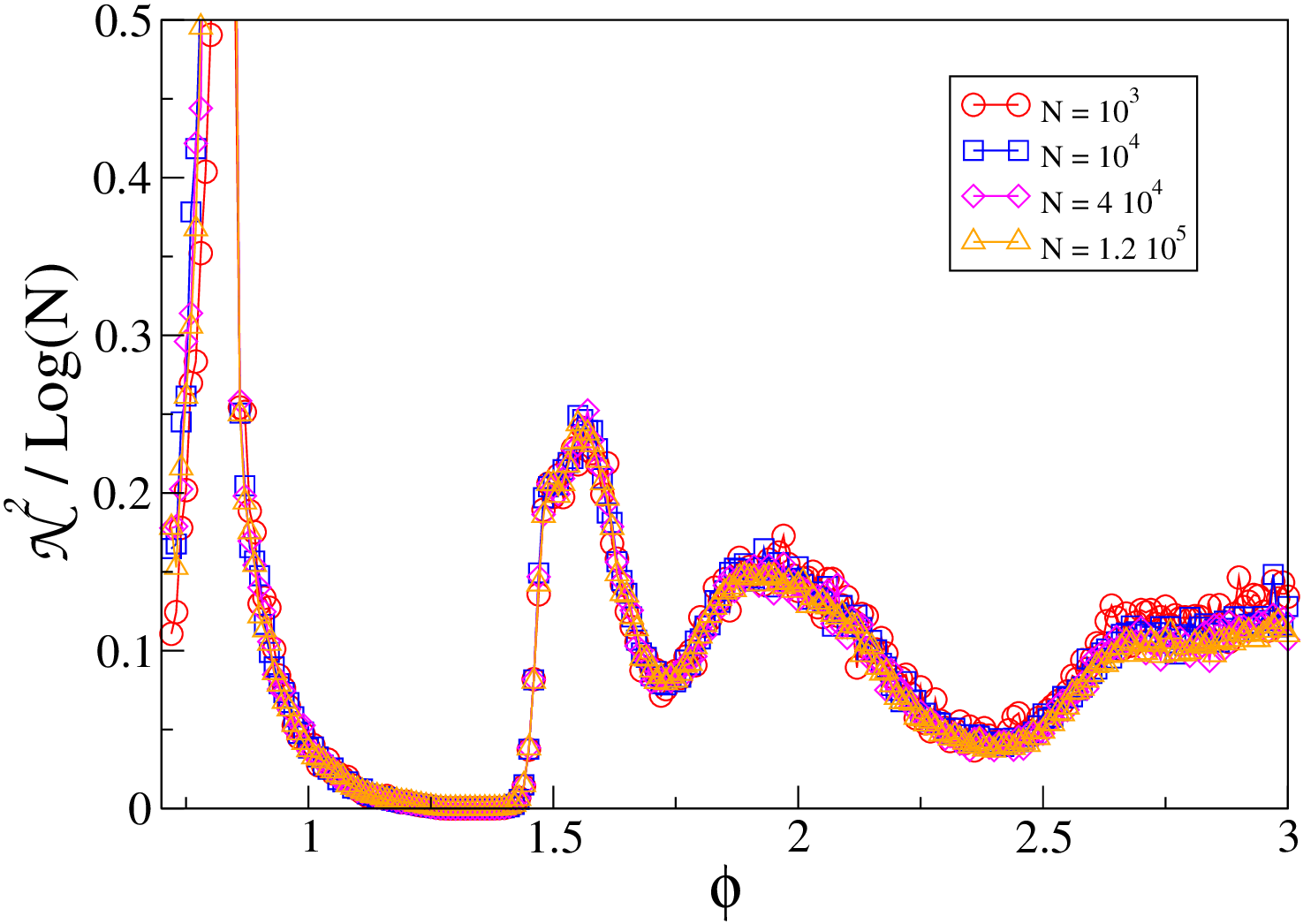}
\caption{
\label{fig:logna}
System size scaling of the non-affine parameter $\na$, suggesting that $\na \propto \log(N)$.}
\end{figure}

Insight into the kind of non-affine rearrangements occurring on compression is obtained
by investigating the system size dependence of $\na$, considering that
$\na$ measures the fluctuations of the non-affine displacement field,
normalized by the number of particles. Accordingly, if the non-affine displacement
originates from independent events each involving $n$ particles, then $\na$ should be 
system size independent for $N \gg n$.
What we actually find, however, is that $\na$ is size dependent for all values of $N$ considered, up to $N = 1.2~10^5$,
and scales as
\begin{equation}
\na \propto \log(N), 
\end{equation}
as illustrated by the data collapse in Fig.~\ref{fig:logna}.

To understand this scaling, we make a connection with a recent study of the
fluctuations of the transverse displacement field in athermal systems under shear~\cite{LemaitreCaroli2009}.
These fluctuations also scale logarithmically with system size. This is rationalized by assuming
that relaxation occurs via localized rearrangements that induce long-range Eshelby--like strain fields.
Indeed, the fluctuations of the Eshelby displacement field ${\bf u} \propto \frac{xy}{r^4} {\bf r}$ of an event located at the origin
scale as ${u^2} \propto \log{L}$~\cite{LemaitreCaroli2009}. This suggests that the dynamics of disordered
systems under quasistatic athermal compression proceeds via a series of pressure avalanches, in the same way that the dynamics 
of disordered systems under quasistatic athermal shear proceeds via a series of shear stress avalanches.

\subsubsection{Avalanches}

\begin{figure}[t!!]
\includegraphics*[scale = 0.31]{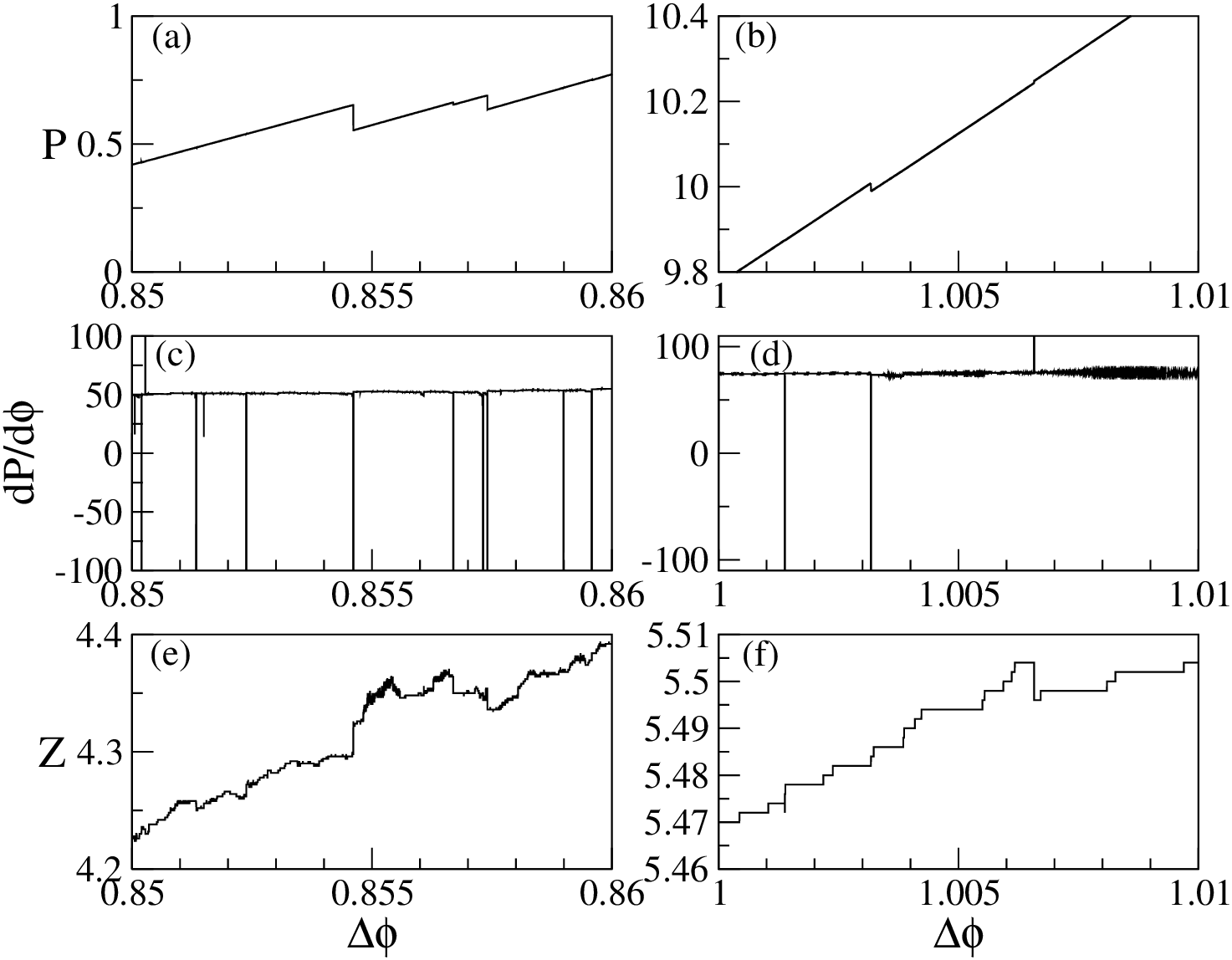}
\caption{
\label{fig:av}
Avalanches induced by a quasistatic compression of the system, starting from $\phi = 0.85$ (left),
and from $\phi = 1.00$ (right). 
The panels illustrate the variation of the pressure (a,b), of its volume fraction derivative (c,d), 
and of the average contact number (e,f).}
\end{figure}
We reveal the avalanches directly by investigating the evolution
of the pressure under quasistatic compression, in an $N = 10^3$ particle system.
We have compressed the system by repeatedly minimizing the energy after small increments
of the volume fraction by $\Delta\phi = 10^{-6}$.
Panels a,b and c,d of Fig.~\ref{fig:av} illustrate the dependence of the pressure and its volume
fraction derivative on the volume fraction, and reveal the presence of
sudden pressure drops corresponding to compression-induced avalanches.
Panels e,f of the figure show the volume fraction evolution of the average contact number $Z$.
Because of the small volume fraction increments we are considering, $Z$ is characterized
by jumps that occur when a contact is created or destroyed. Most frequently,
avalanches are correlated with a decrease of the average number of contacts.

Under quasistatic athermal shear, there is a typical shear strain separation between successive avalanches, $\Delta \gamma_{av}$,
that depends on the system size, on the density, and on the potential. Similarly, one may expect here the presence of a
typical volume fraction separation between successive compression induced avalanches, $\Delta \phi_{av}$.
When on compressing the system one reaches compressions $\Delta \phi > \Delta \phi_{av}$, then avalanches
occur and (typically~\cite{PCS2}) decrease the pressure. This could explain the crossover in the
dependence of the pressure on the volume fraction observed in Fig.~\ref{fig:DP}. 
More work is needed to extract the volume
fraction dependence of $\Delta \phi_{av}$,
for which one would expect the scaling with system size $\Delta \phi_{av} \propto \frac{1}{N} \Delta \phi_{av}^0$.
The intensive coefficient $\Delta \phi_{av}^0$ may then lead to the identification of a new 
avalanche-related length scale, $\xi \propto \left(\Delta \phi_{av}^0\right)^{-1/d}$ in $d$ spatial dimensions.

\section{Conclusions}
We have discussed some of the features of high order jamming crossovers
in systems of particles interacting via finite ranged repulsive potentials,
focusing on the volume fraction dependence of the degree of non-affinity of the system.
To this end, we introduced two different measures of non-affinity, 
one obtained by comparing the actual bulk modulus with that predicted
in the affine approximation, the other directly measuring the non-affine
displacement field. These two quantities show clearly that the jamming crossovers induce
an increase in the non affine response of the system, in much the same way as at the standard jamming transition.

We have rationalized the logarithmic system size scaling of the non affine parameter $\na$
by assuming that upon compression jammed systems evolve via avalanches; these consist of a collection of localized rearrangements that
induce long-ranged Eshelby displacement fields. We have directly investigated
these avalanches, observing pressure drops under compression. 

Our work suggests as an interesting future topic the investigation of the volume
fraction dependence of the typical volume fraction increment $\Delta \phi_{av}$
required to trigger an avalanche, as this may be related to a
new length scale characterizing jammed packings.

\begin{theacknowledgments}
MPC thanks the Dept.\ of Mathematics, King's College London for hospitality,
and MIUR-FIRB RBFR081IUK for financial support.
\end{theacknowledgments}



\bibliographystyle{aipproc}   


\IfFileExists{\jobname.bbl}{}
 {\typeout{}
  \typeout{******************************************}
  \typeout{** Please run "bibtex \jobname" to optain}
  \typeout{** the bibliography and then re-run LaTeX}
  \typeout{** twice to fix the references!}
  \typeout{******************************************}
  \typeout{}
 }

\end{document}